\documentclass[twocolumn,showpacs,preprintnumbers,amsmath,amssymb,prl,superscriptaddress]{revtex4}
\usepackage{graphicx}
\usepackage{dcolumn}
\usepackage{bm}

\newcommand{\fmn}[2]{\mbox{${\textstyle \frac{#1}{#2}}$}}
\newcommand{\dd}{\mbox{\text{d}}}

\begin{document}

\title{Precision study of the $\eta\,^{3}$He system using
the $dp{\to}\,^{3}$He$\eta$ reaction}
\author{T.\,Mersmann}
\affiliation{Institut f\"ur Kernphysik, Universit\"at
M\"unster, 48149 M\"unster, Germany}%
\author{A.\,Khoukaz}\email[E-mail: ]{khoukaz@uni-muenster.de}
\affiliation{Institut f\"ur Kernphysik, Universit\"at
M\"unster, 48149 M\"unster, Germany}%
\author{M.\,B\"{u}scher}
\affiliation{Institut f\"ur Kernphysik, Forschungszentrum
J\"ulich, 52425 J\"ulich, Germany}
\author{D.\,Chiladze}
\affiliation{Institut f\"ur Kernphysik, Forschungszentrum
J\"ulich, 52425 J\"ulich, Germany}
%
\author{S.\,Dymov}
\affiliation{Laboratory of Nuclear Problems, JINR, 141980 Dubna, Russia}
\author{M.\,Hartmann}
\affiliation{Institut f\"ur Kernphysik, Forschungszentrum
J\"ulich, 52425 J\"ulich, Germany}
\author{V.\,Hejny}
\affiliation{Institut f\"ur Kernphysik, Forschungszentrum
J\"ulich, 52425 J\"ulich, Germany}
\author{A.\,Kacharava}
%
\affiliation{Physikalisches Institut II, Universit{\"a}t
Erlangen-N{\"u}rnberg, 91058 Erlangen, Germany}%
\author{I.\,Keshelashvili}
\affiliation{Institut f\"ur Kernphysik, Forschungszentrum
J\"ulich, 52425 J\"ulich, Germany}
\author{P.\,Kulessa}
\affiliation{Institute of Nuclear Physics, Cracow University, 31342 Cracow, Poland}
\author{Y.\,Maeda}
\affiliation{Research Center for Nuclear Physics, Osaka University, Ibaraki, Osaka 567-0047, Japan}
\author{M.\,Mielke}%
\affiliation{Institut f\"ur Kernphysik, Universit\"at
M\"unster, 48149 M\"unster, Germany}%
\author{S.\,Mikirtychiants}
\affiliation{High Energy Physics Department, Petersburg Nuclear Physics Institute, 188350 Gatchina, Russia}
\author{H.\,Ohm}
\affiliation{Institut f\"ur Kernphysik, Forschungszentrum
J\"ulich, 52425 J\"ulich, Germany}
\author{M.\,Papenbrock}%
\affiliation{Institut f\"ur Kernphysik, Universit\"at
M\"unster, 48149 M\"unster, Germany}%
\author{D.\,Prasuhn}
\affiliation{Institut f\"ur Kernphysik, Forschungszentrum
J\"ulich, 52425 J\"ulich, Germany}
\author{F.\,Rathmann}
\affiliation{Institut f\"ur Kernphysik, Forschungszentrum
J\"ulich, 52425 J\"ulich, Germany}
\author{T.\,Rausmann}%
\affiliation{Institut f\"ur Kernphysik, Universit\"at
M\"unster, 48149 M\"unster, Germany}%
\author{R.\,Schleichert}
\affiliation{Institut f\"ur Kernphysik, Forschungszentrum
J\"ulich, 52425 J\"ulich, Germany}
\author{V.\,Serdyuk}
\affiliation{Laboratory of Nuclear Problems, JINR, 141980 Dubna,
Russia}
\author{H.-J.\,Stein}
\affiliation{Institut f\"ur Kernphysik, Forschungszentrum
J\"ulich, 52425 J\"ulich, Germany}
\author{H.\,Str\"oher}
\affiliation{Institut f\"ur Kernphysik, Forschungszentrum
J\"ulich, 52425 J\"ulich, Germany}
\author{A.\,T\"{a}schner}
\affiliation{Institut f\"ur Kernphysik, Universit\"at
M\"unster, 48149 M\"unster, Germany}%
\author{Yu.\,Valdau}
\affiliation{High Energy Physics Department, Petersburg Nuclear Physics Institute, 188350 Gatchina, Russia}
\author{C.\,Wilkin}
\affiliation{Physics and Astronomy Department, UCL, Gower Street,
London WC1E 6BT, UK}%
\author{A.\,Wro\'{n}ska}
\affiliation{Institute of Nuclear Physics, Cracow University, 31342 Cracow, Poland}
%
%
\date{\today}

\begin{abstract}
The differential and total cross sections for the
$dp\to\,^{3}\textrm{He}\,\eta$ reaction have been measured in a
high precision high statistics COSY--ANKE experiment near threshold
using a continuous beam energy ramp up to an excess energy $Q$ of
11.3\,MeV with essentially 100\% acceptance. The kinematics
allowed the mean value of $Q$ to be determined to about 9\,keV.
Evidence is found for the effects of higher partial waves for
$Q\agt 4\,$MeV. The very rapid rise of the total cross section to
its maximum value within $0.5\,$MeV of threshold implies a very
large $\eta\,^3$He scattering length and hence the presence of a
quasi--bound state extremely close to threshold.
\end{abstract}

 \maketitle

The low energy $\eta\,^{3}$He system  has been investigated in
both the $dp\:(pd)\to\,^{3}\textrm{He}\,\eta$
reactions~\cite{Berger,Mayer} as well as in
photoproduction~\cite{Pfeiffer}. The anomalous energy dependence
found there suggests that the strong $\eta\,^{3}$He final state
interaction (\textit{fsi}) might lead to the formation of a new
state of matter in the form of an $\eta\,^{3}$He quasi--bound
state~\cite{Wilkin} for nuclei much lighter than originally
postulated~\cite{Liu}. This question is far from being settled and
further high quality data are required to inform the debate.

The SPESII measurements~\cite{Mayer} show that the absolute square
of the production amplitude $|f|^2$ falls by a factor of three
between threshold and an excess energy
$Q=\sqrt{s}-m_{\eta}-m_{^3{\text{He}}}$ of about 6\,MeV and the
conclusion drawn was that the $\eta\,^{3}$He scattering length $a$
is very large. However, in order to distinguish between the
effects of the real and imaginary parts of $a$, one needs data
with a good knowledge of the absolute value of $Q$ and with a very
small energy spread. These conditions are hard to meet in the
interesting near--threshold region when using a liquid hydrogen
target and an extracted proton beam because of the energy losses
in the target. On the other hand, they can be overcome by using a
thin windowless gas target placed inside a storage ring, and we
have taken advantage of these features in a measurement of the
$dp\to\,^{3}\textrm{He}\,\eta$ reaction near threshold.

The experiment was performed with a hydrogen cluster--jet
target~\cite{target} using the ANKE spectrometer~\cite{ANKE}
placed at an internal station of the COoler SYnchrotron
COSY--J{\"u}lich. During each of the beam cycles of 277 seconds,
the deuteron beam energy was ramped slowly and linearly in time,
from an excess energy of $Q = -5.05\,$MeV to $Q = +11.33\,$MeV.
The $^3$He produced were detected in the ANKE forward detection
system, which consists of two multi-wire proportional chambers,
one drift chamber and three layers of scintillation hodoscopes.
The geometrical acceptance for the $^3$He of interest was $\sim
100$\%, so that systematic uncertainties from acceptance
corrections are negligible. The tracks of charged particles could
be traced back through the precisely known magnetic field to the
known interaction point, leading to a momentum reconstruction for
registered particles. The luminosity required to determine cross
sections was found by simultaneously measuring $dp$ elastic
scattering, with the scattered deuterons being registered in the
forward detector and the proton reconstructed from the missing
mass. Data in this region show that $\dd\sigma/\dd{t}$ changes
little with beam energy, though the dependence on the
four--momentum transfer in our available range of
$0.08<|t|<0.20\,$(GeV/c)$^2$ is very strong~\cite{dpelastic}. It
is therefore the systematic uncertainty in the measurement of the
deuteron scattering angle that dominates the error of about
$\pm15$\% in the luminosity determination. However, it must be
stressed that the relative luminosity over the ramp is known much
better and its effects are included in the point--by--point
errors.

\begin{figure}[hbt]
\includegraphics[width=8.5cm]{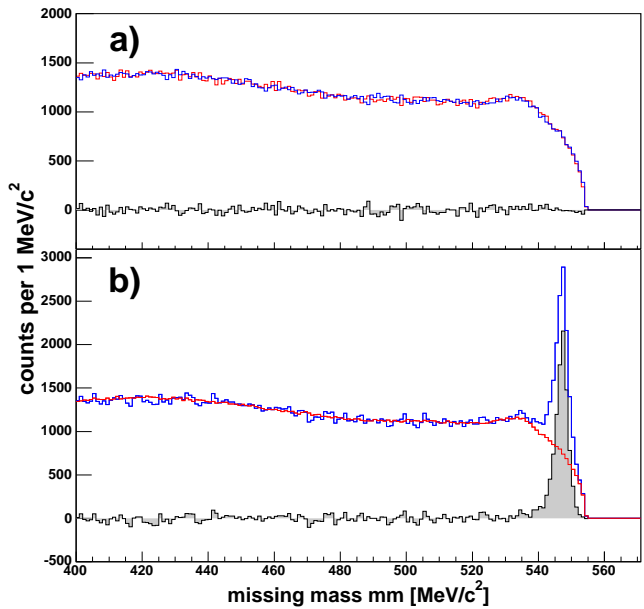}%
\caption{\label{fig:backg} (Color online) Missing--mass
distributions of detected $^3$He nuclei. (a) Measurements at the
subthreshold energies of $Q=-(4.75 \pm 0.30)\,$MeV and
$Q=-(0.60\pm0.30)\,$MeV were analyzed as if they were both taken
at $Q = +6.95\,$MeV. After correcting for luminosity, the two
distributions look identical and the difference shown is
consistent with zero. (b) The $Q=6.95\,$MeV distribution is
compared with the background description derived from all the
subthreshold data. The difference spectrum shows only the very
clean $\eta$ peak (shaded) with
$\sigma_{\text{mm}}=2.0\,\textrm{MeV/c}^2$.}
\end{figure}

The $^3$He were selected by the $\Delta E/E$ method, with the
$\eta$ meson being subsequently identified through a peak in the
missing--mass distribution~\cite{Ola}. The distribution for the
interval $Q = (6.95 \pm 0.12)\,$MeV is presented in
Fig.~\ref{fig:backg}b. In order to extract the total number of
$\eta$ events, the background below the $\eta$ peak, originating
mainly from multi--pion production, has to be subtracted. For this
purpose, measurements performed below the $\eta$ threshold were
analyzed as if they were made above~\cite{Ola}. The reliability of
this approach is made clear in Fig.~\ref{fig:backg}a, where events
obtained at $Q=-(4.75 \pm 0.30)\,$MeV and $Q=-(0.60\pm0.30)\,$MeV
were both analyzed assuming a value of $Q = +6.95\,$MeV. After
correcting for luminosity, the two distributions coincide
perfectly and their difference is consistent with zero. This
technique is applied above threshold in Fig.~\ref{fig:backg}b,
where the distribution corresponding to all the measurements with
$Q<-0.3\,$MeV has been subtracted from the $Q=6.95\,$MeV
missing--mass spectrum. All that remains in the difference
spectrum is an $\eta$ meson peak with a width of
$\sigma_{\text{mm}}=2.0\,\textrm{MeV/c}^2$ sitting on a
vanishingly small background.

\begin{figure}[hbt]
\includegraphics[width=8.5cm]{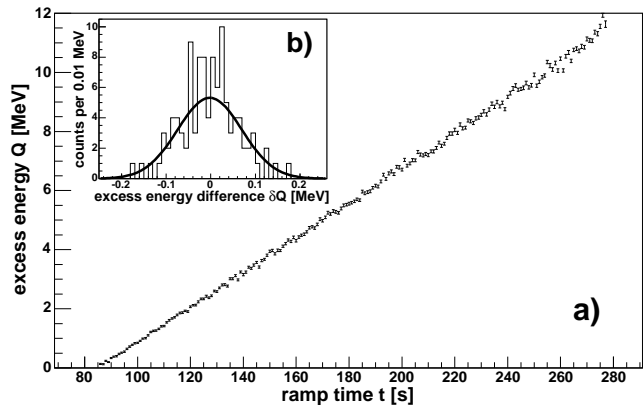}%
\caption{\label{fig:excess} (a) Reconstructed excess energy $Q$ as
a function of the timing information of the linearly ramped beam.
(b) Distribution of the deviations from a linear fit to the time
dependence for different bins in excess energy.}
\end{figure}

The excess energy at a particular time during the ramp was
reconstructed by studying the size of the $^3$He momentum locus in
the c.m.\ frame and comparing it with analytic formulae and
simulations. The consistency of the excess energy determination
with the expected linear variation of the beam energy with ramp
time is demonstrated in Fig.~\ref{fig:excess}a. The deviations
from the linear fit presented in panel \ref{fig:excess}b are
consistent with a statistical distribution of width
$\sigma_{\delta Q}=(72\pm11)\,$keV. The mean value could therefore
be determined to 9\,keV and this becomes even more precise as one
approaches threshold. Since, due to the uncertainty in the orbit
length, the absolute beam momentum is not known to better than
about 3\,MeV/c, this excellent measurement of $Q$ cannot be
translated into an accurate value of the  $\eta$ mass.

\begin{figure}[hbt]
\includegraphics[width=8.5cm]{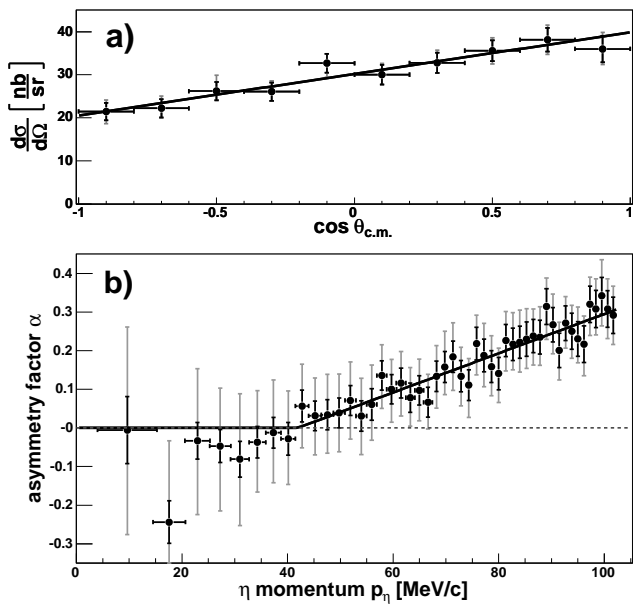}%
\caption{\label{fig:angul} (a) $dp\to\,^{3}\textrm{He}\,\eta$
differential cross section at an excess energy of
$Q=(10.27\pm0.12)\,$MeV as a function of the cosine of the $^3$He
c.m.\ emission angle. (b) Asymmetry parameter $\alpha$ for
different values of the $\eta$ c.m.\ momentum $p_{\eta}$. The bin
widths and point--to--point statistical errors are shown bold;
correlated systematic uncertainties are shown with feint lines.}
\end{figure}

In order to determine the differential cross section for each
excess energy bin, missing--mass distributions were analyzed for
different $^3$He c.m.\ production angles $\theta_{\text{c.m.}}$ in
a similar manner to that shown in Fig.~\ref{fig:backg}. The
asymmetry in the angular distribution of Fig.~\ref{fig:angul}a
implies that there are higher partial wave contributions even in
this very near--threshold region. Defining an asymmetry parameter
$\alpha$ through $(\dd\sigma/\dd\Omega)_{\text{c.m.}}=
\sigma_{\text{tot}}\,(1+\alpha\cos\theta_{\text{c.m.}})/4\pi$, it
is seen from Fig.~\ref{fig:angul}b that above an $\eta$ c.m.\
momentum $p_{\eta}$ of 40\,MeV/c ($Q=1.7\,$MeV), $\alpha$ is
positive and increases monotonically with  $p_{\eta}$ but with a
magnitude much larger than that found at SPESII~\cite{Mayer}. At
low momentum, both data sets show a tendency for $\alpha$ to go
negative but the systematic uncertainties here are large. The
slope of the cross section at $\cos\theta_{\text{c.m.}}=0$ has the
same sign as that found at higher excess energies, though the data
there do not remain linear over the whole angular
range~\cite{GEM,Jozef,Adam}.

\begin{figure*}[hbt]
\includegraphics[width=16.0cm]{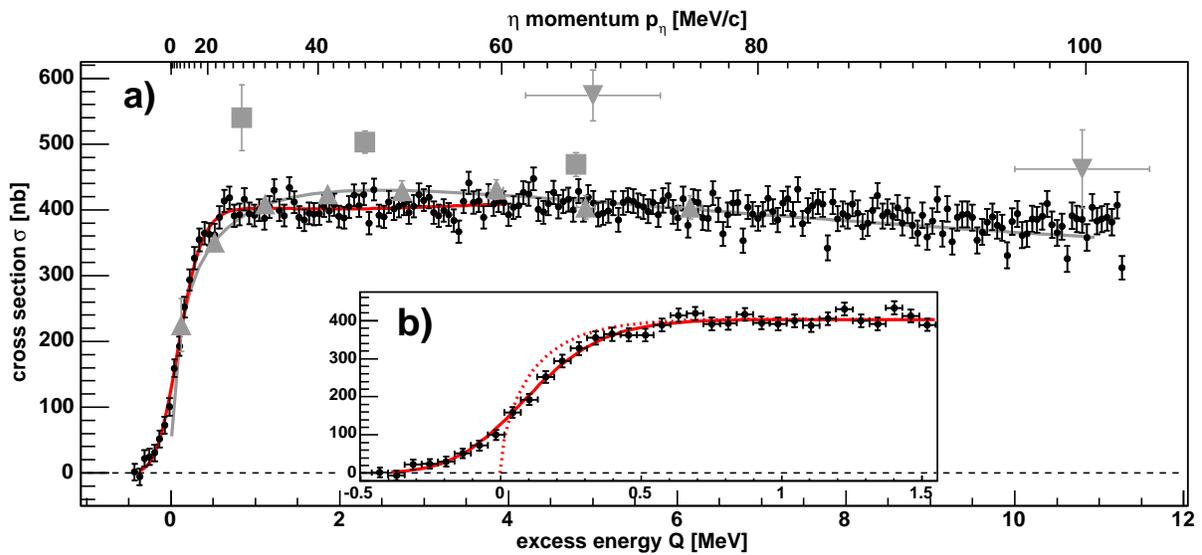}%
\caption{\label{fig:cross} (Color online) Comparison of the
extracted total cross sections (circles) with previous data drawn
in gray: Ref.~\cite{Berger} (squares), Ref.~\cite{Mayer}
(triangles), and Ref.~\cite{Adam} (inverted triangles). The fit to
our results for $Q<4\,$MeV corresponds to the parameters given in
Table~\ref{tab:fit}. The gray curve is the SPESII fit to their own
data~\cite{Mayer}. Our near--threshold data and fitted curve are
shown in the inset, while the dotted curve is the result to be
expected without the 180\,keV smearing in $Q$.}
\end{figure*}

The $dp\to\,^{3}\textrm{He}\,\eta$ total cross sections obtained
at 195 bins in excess energy $Q$ are displayed in
Fig.~\ref{fig:cross}. The minimal relative systematic errors
resulting from the measurement of the excitation function in a
single experiment form a robust data set for any phenomenological
analysis. Our data are broadly compatible with those of
SPESII~\cite{Mayer} and any global difference is within our
overall normalization uncertainty. However, in contrast to our
data presented in Fig.~\ref{fig:cross}b, the SPESII results do not
define firmly the energy dependence in the near--threshold region.
The total cross section reaches its maximum value within 0.5\,MeV
of threshold and hardly decreases after that. This behavior is in
complete contrast to phase--space expectations and indicates a
very strong final state interaction~\cite{Wilkin}.

The angular average of the amplitude--squared is derived from the
total cross section $\sigma_{\text{tot}}$ through
\begin{equation}
\label{fsq}%
\overline{|f|^2}=\frac{1}{4\pi}\,\frac{p_d}{p_\eta}\,\sigma_{\text{tot}},
\end{equation}
where $p_d$ is the initial c.m.\ momentum. In the
presence of a strong final state interaction, the $s$--wave
amplitude $f_s$ can be approximated in terms of the complex
$\eta\,^3$He scattering length ($a$) and effective range
($r_0$)
\begin{equation}
f_s = f_B\left/\left(1-iap_\eta +\fmn{1}{2} a r_0
p_\eta^2\right)\right., \label{eq:fsi2}
\end{equation}
where $f_B$ is assumed to be slowly varying.

The shape of the $\eta$ production below the nominal threshold
shown in Fig.~\ref{fig:cross} is a very sensitive measure of the
momentum width of the COSY beam, and this resolution has to be
taken into account in any phenomenological analysis. The
excitation function $\sigma_{\text{exp}}(Q)$ visible in the
experiment is the convolution of the true one
$\sigma_{\text{true}}(Q)$ with a smearing function $w_g(Q)$, taken
to be Gaussian, which is then grouped in finite bins of
$Q\pm\Delta Q/2$:
\begin{equation}
\sigma_{\text{exp}}(Q)=\frac{1}{\Delta Q} \int\limits_{Q-\Delta
Q/2}^{Q+\Delta Q/2} \dd{Q}_1
\int\limits_{-\infty}^{+\infty}\dd{Q}_2 \, w_{g}(Q_1,Q_2)\,
\sigma_{\text{true}}(Q_2), \label{eq:wq}
\end{equation}
where the luminosity is assumed constant over each bin of a
typical $\Delta Q=0.06\,$MeV width. The fitting process shows that
the resolution in $Q$ is $\sigma_Q=(180\pm15)\,$keV. If this
arises purely from the momentum spread of the beam it would
correspond to $\left(\delta p/p\right)_{\,\text{beam}}\approx
2.3\times 10^{-4}$. The effect of the smearing in $Q$ is
illustrated in Fig.~\ref{fig:cross}b, where the unsmeared
parametrization, $\sigma_{\text{true}}$, is shown by the dotted
curve.

The gray line in Fig.~\ref{fig:cross}a shows the published fit of
Eq.~(\ref{eq:fsi2}) to the SPESII results~\cite{Mayer} without the
effective range term and without smearing over the effective beam
energy. Although this parametrizes these data very well, it
underestimates the rapid rise from threshold in our more extensive
data set. In fact any fit of Eq.~(\ref{eq:fsi2}), smeared
according to Eq.~(\ref{eq:wq}), to the present data that neglects
the effective range $r_0$ fails to satisfy simultaneously the data
in the proximity of threshold ($Q\leq 1\,$MeV) and at the higher
energies ($Q\ge 3\,$MeV). This effect becomes visible here for the
first time because of the quality and extent of the data. This
lack of a successful fit with the simple scattering length formula
might be caused by contributions from higher partial waves or by
effective range effects.

The presence of higher partial waves is obvious from the angular
distributions of Fig.~\ref{fig:angul} but, if the asymmetry is due
to $s$--$p$ interference, this could be explained by a mere 3\%
$p$--wave contribution. The identification of the higher partial
waves requires experiments with a polarized deuteron beam which
will be performed~\cite{Rausmann}.

The inclusion of the effective range term results in a
significantly better description of the data, as can be seen from
the solid line in Fig.~\ref{fig:cross}a. To minimize the effects
of the higher partial waves, only the ANKE data up to an excess
energy of $Q = 4\,$MeV have been considered in the fit, whose
results are summarized in Table~\ref{tab:fit}.
\begin{table}[htb]
\begin{ruledtabular}
\begin{tabular}{cc}
Scattering length (fm) & Effective range (fm)
 \\ \hline
 $\textrm{Re}(a)=\hspace{1mm}11.6\pm1.4^{+0.3}_{-0.4}$
 & $\textrm{Re}(r_0)=-2.0\pm1.1^{+0.3}_{-0.1}$ \\
 $\textrm{Im}(a)=-4.1\pm7.0^{+1.9}_{-1.2}$
 & $\textrm{Im}(r_0)=\phantom{-}2.8\pm1.5^{+0.2}_{-0.3}$ \\
\end{tabular}
\end{ruledtabular}
\caption{\label{tab:fit}Real and imaginary parts of the scattering
length $a$ and effective range $r_0$ of the $\eta\,^3$He system,
derived by fitting the ANKE data for $Q<4\,$MeV with
Eqs.~(\ref{fsq}) and (\ref{eq:fsi2}) smeared over the beam
resolution using Eq.~(\ref{eq:wq}). The first error is statistical
and the second systematic, including effects arising from the
choice of fitting range. It is important to note that the data are
not sensitive to the overall sign of the real parts and also that
the errors are strongly correlated.}
\end{table}

In order to affect the cross section variation over a scale of
less than 1\,MeV, there must be a pole of the production amplitude
in the complex plane that is typically only 1\,MeV away from
$Q=0$. Using the parameters of Table~\ref{tab:fit} in
Eq.~(\ref{eq:fsi2}), the nearby pole position is found to be
$p_{\eta}=[(4.1\pm12.4\pm1.6)\pm i(17.6\pm6.9\pm1.0)]$\,MeV/c. In
the energy plane, the pole is at $Q_0=p_{\eta}^2/2m_{\text{red}} =
[(-0.20\pm0.38\pm0.04)\pm i(0.16\pm0.52\pm0.06)]\,$MeV, where
$m_{\text{red}}$ is the $\eta\,^3$He reduced mass. Though the
errors are strongly correlated, as they are for the scattering
length and effective range, the pole position is stable to within
a fraction of an MeV.

In summary, we have performed measurements of the differential and
total cross sections for the $dp\,\rightarrow\,^3\text{He}\,\eta$
reaction near threshold where the spectrometer acceptance is close
to 100\%. The use of a beam whose energy varies linearly with time
ensured that point--to--point systematic errors were under
control. It also allowed us to determine the mean value of the
excess energy with unparalleled accuracy. It was shown that the
large physics background could be eliminated essentially
completely through the subtraction of data taken below threshold.
Although there is a 15\% uncertainty in the luminosity, and hence
in the values of the cross sections, this is a global feature that
affects all our data in the same way and so will not change any of
our principal conclusions.

It is remarkable that already for $Q\agt 4\,$MeV the angular
distributions are no longer isotropic and this must be an
important clue to the dynamics. Effects of $p$ waves might become
clearer when data are available on the angular dependence of the
deuteron analyzing powers~\cite{Rausmann}. Extra information will
also become available from an unpolarized COSY-11
measurement~\cite{Smyrski}.

The consistent set of total cross section measurements with high
statistics at closely spaced values of $Q$ should allow
theoretical models to be tested in a rigorous manner. The very
rapid rise and levelling--off indicates the existence of a pole in
the production amplitude within one MeV of $Q=0$. Simple fits,
using an effective range approximation for the final state
interaction, suggest that the scattering length has an enormous
real part that largely masks any effects arising from the
imaginary part. The steep variation of $\overline{|f|^2}$ with
$p_{\eta}$ may bring the results closer to those of
photoproduction of the $\eta\,^3$He state~\cite{Pfeiffer}.

\begin{acknowledgments}
Our experiment was only possible because of the high quality of
the ramped deuteron beam and for this we are indebted to the COSY
accelerator crew. We would also like to thank Ch.\ Hanhart for
many valuable discussions. The support from FFE grants of the
J{\"u}lich Research Center is gratefully acknowledged.
\end{acknowledgments}


\end{document}